\documentclass[12pt,english,nofootinbib,preprint]{revtex4-1}
\usepackage{hyperref}
\usepackage{setspace}
\usepackage{mathptmx}
\usepackage{graphicx}

\usepackage[T1]{fontenc}
\usepackage[latin9]{inputenc}
\usepackage[letterpaper]{geometry}
\usepackage{booktabs}
\usepackage{amsmath}
\usepackage{amssymb}
\usepackage{cancel}
\usepackage{babel}

\newcommand {\be}{\begin{equation}}
\newcommand {\ee} {\end{equation}}
\newcommand {\bea}{\begin{eqnarray}}
\newcommand {\eea} {\end{eqnarray}}

\begin{document}

\title{Metastates and Replica Symmetry Breaking}
\author{C.M.~Newman}
\affiliation{ Courant Institute of Mathematical Sciences, New York University, New York, NY 10012 USA }
\affiliation{NYU-ECNU Institute of Mathematical Sciences at NYU Shanghai, 3663 Zhongshan Road North, Shanghai, 200062, China}
\author{N.~Read}
\affiliation{ Department of Physics, Yale University, P.O. Box 208120, New Haven, Connecticut 06520-8120, USA}
\affiliation{Department of Applied Physics, Yale University, P.O. Box 208284, New Haven, Connecticut 06520-8284, USA}
\author{D.L.~Stein}
\affiliation{Department of Physics and Courant Institute of Mathematical Sciences, New York University, New York, NY 10012 USA}
\affiliation{NYU-ECNU Institutes of Physics and Mathematical Sciences at NYU Shanghai, 3663 Zhongshan Road North, Shanghai, 200062, China}
\affiliation{Santa Fe Institute, 1399 Hyde Park Rd., Santa Fe, NM USA 87501}

\begin{abstract}
In this review we define and discuss metastates, mathematical tools with general applicability to thermodynamic systems which are particularly useful when 
working with disordered or inhomogeneous short-range systems.  In an infinite such system there may be many competing thermodynamic states, 
which can lead to the absence of a straightforward thermodynamic limit of local correlation functions. A metastate is a probability measure on the infinite-volume 
thermodynamic states that restores the connection between those states and the Gibbs states observed in finite volumes. 
After introducing the basic metastates and discussing their properties, we present possible scenarios for the spin-glass phase and discuss what the metastate 
approach reveals about how replica symmetry breaking would manifest itself in finite-dimensional short-range spin glasses. 
%We define and discuss the notion of metastate, a mathematical tool with general applicability to thermodynamic systems but which is particularly 
%useful when
%working with disordered or inhomogeneous systems.  Such systems may have many competing thermodynamic states, which can lead to the 
%absence of a 
%straightforward thermodynamic limit of local correlations. The metastate is a probability measure on the thermodynamic states themselves, allowing 
%one to restore in such cases the
%connection between the thermodynamics of infinite volumes and Gibbs states observed in finite volumes. After introducing the metastate and 
%discussing its properties, we present possible scenarios for the spin glass phase and discuss what the metastate approach reveals about how replica 
%symmetry breaking would manifest itself in finite-dimensional spin glasses.
\end{abstract}

\maketitle

\section{Introduction}
\label{sec:intro}

The presence of quenched disorder in condensed matter systems creates special challenges for statistical mechanics, designed primarily to deal with
thermal disorder.  A proper statistical mechanical treatment of a system whose Hamiltonian contains quenched random variables requires averaging the free energy,
rather than the partition function, over these variables. This led to the so-called ``replica trick'', first used in the context of spin glasses by Edwards and 
Anderson~\cite{EA75} (EA).
%(an earlier use of the replica trick, following a suggestion of Mark Kac, was by Sam Edwards~\cite{Edwards71} in studies of rubber polymers). 
Sherrington and Kirkpatrick~(SK)~\cite{SK75} applied the replica trick to an infinite-range spin glass model, using the ``spin freezing'' order parameter proposed 
in~\cite{EA75}, but found the solution to be thermodynamically unstable at low temperature. The solution, found by Giorgio Parisi in 1979~\cite{Parisi79}, 
required an exotic and non-intuitive notion of ``replica symmetry breaking'' (RSB); it was a stunning result that led (in Phil Anderson's 
words~\cite{AndersonPT6}) to a ``cornucopia'' of applications, to which this volume is dedicated.

A related problem, on which the effectiveness of theoretical statistical mechanics depends, is the existence of a thermodynamic limit for the state 
in which a system resides. When encountering the subject for the first time a student typically learns that for a macroscopic system comprising 
$N\sim 10^{24}-10^{25}$ degrees of freedom with short-range interactions, the effects of surfaces on bulk
thermodynamic properties vanishes for large system size, so that a straightforward $N\to\infty$ limit describes the thermodynamics of a large but finite sample. 
This approach works well in describing the condensed phase for systems without quenched disorder (so long as an appropriate order parameter is identified), 
as well as for some systems {\it with\/} quenched disorder. But when quenched disorder is present, its success cannot be guaranteed, particularly if the 
low-temperature phase consists of many pure states unrelated by any simple symmetry transformation.

This is not merely a theoretical issue. Consider for example a dilute magnetic alloy such as CuMn, which for a range of Mn concentrations displays spin glass 
behavior, and two labs which prepare their samples under similar conditions. The locations of the magnetic impurity Mn atoms within the Cu lattice will differ 
in the two samples, leading to a different distribution of ferromagnetic and antiferromagnetic couplings; in more formal language, the two samples have 
different realizations of the spin-spin couplings. It now becomes crucial to understand what effect the difference in coupling realizations will have on the system 
thermodynamics: if there is too much dependence then universal behavior would be absent, i.e., different samples would behave differently.

Fortunately, it can be shown that measurable {\it global\/} properties (i.e., quantities which do not depend on the configuration of any finite subset 
of an infinite system) do {\it not\/} in general depend on coupling realization:\footnote{An important exception is the spin or edge overlap between two states 
in a system with RSB; the fact that this global quantity is ``non-self-averaging'' was one of the surprises of RSB. Non-self-averaging 
in the context of short-range models is discussed in Sect.~\ref{subsubsec:NSA}.} these global quantities include energy/free energy per spin, magnetization 
per spin, transition temperature, and so on. Thermodynamic {\it states\/}, however, are local: they can be considered as enumerations of all $k$-spin 
correlation functions, where $k=1,2,3,\ldots$. As such they are exquisitely sensitive to coupling realization, and the usual straightforward approach of taking 
a simple thermodynamic limit can break down. As we will see, this is especially important when dealing with systems in which the ordering at low temperature 
is described by RSB.

In what follows we will discuss in further detail the breakdown in certain cases of the straightforward thermodynamic limit (i.e., one not conditioned on the 
quenched disorder), and present the concept of a metastate as a statistical-mechanical tool designed to handle this.  We will show how metastates are useful 
for describing the properties of large finite volumes in the absence of a straightforward thermodynamic limit, and investigate some of their uses and applications. 
Given that the volume in which this contribution appears is dedicated to the many applications of RSB, we will focus our attention on how the 
metastate is especially useful when working with systems in which the low-temperature ordering is or may be described by RSB, such as EA spin glasses.

%%%%%%%%%%%%%%%%%%%%%%%%%%%%%%%
\section{Chaotic size dependence and the thermodynamic limit}
\label{sec:csd}

We will be primarily interested in the EA Ising spin glass~\cite{EA75} in zero field. Its (infinite-volume) Hamiltonian is given by
\begin{equation}
\label{eq:EA}
H_J(\sigma)=-\sum_{\{{\bf x},{\bf y}\}} J_{{\bf {\bf xy}}} \sigma_{\bf x}\sigma_{\bf y} 
\end{equation}
where ${\bf x}$, ${\bf y}\ldots \in\mathbb{Z}^d$ are sites in the $d$-dimensional hypercubic lattice $\mathbb{Z}^d$, $\sigma_{\bf x}=\pm 1$ 
is an Ising spin at site ${\bf x}$ and $\{ {\bf x},{\bf y}\}$ denotes an ``edge'' in the set $\mathbb{E}^d$ of nearest-neighbor pairs. The couplings (or bonds) 
$J_{{\bf {\bf xy}}}=J_{{\bf yx}}$ are independent, identically-distributed continuous random variables chosen from a distribution $\nu(dJ_{{\bf {\bf xy}}})$, 
with random variable $J_{{\bf xy}}$  assigned to the edge $\{ {\bf x},{\bf y}\}$.  For simplicity we will assume that $\nu$ is supported on the entire real line, 
is distributed symmetrically about zero, and has finite variance; e.g., a Gaussian with mean zero and variance one.  We denote by $J$ a particular realization 
of the couplings.

Theoretically and numerically (not to mention experimentally!) one necessarily works with finite volumes. Let $\Lambda_L\subset\mathbb{Z}^d$ denote 
a cube of side $L$ centered at the origin, $\vert\Lambda_L\vert$ denote its volume, i.e., the number of spins contained within $\Lambda_L$, and 
$\partial\Lambda_L$ its boundary. One then considers $H_{J,L}(\sigma)$, the Hamiltonian~(\ref{eq:EA}) restricted to $\Lambda_L$ with a specified 
boundary condition on $\partial\Lambda_L$, usually taken to be periodic, which among other advantages preserves the spin-flip symmetry of the Hamiltonian. 
The thermodynamics within $\Lambda_L$ is described by the finite-volume Gibbs distribution $\rho^{(L)}$: for any (well-behaved) function $f(\sigma)$ of 
some subset of the spins within $\Lambda_L$, its expectation at inverse temperature $\beta=1/T$ is given by
\begin{equation}
\label{eq:Gibbs}
\langle f(\sigma)\rangle_{\rho^{(L)}}=\sum_{\{\sigma_L\}} f(\sigma)e^{-\beta H_{J,L}(\sigma)}/\sum_{\{\sigma_L\}}e^{-\beta H_{J,L}(\sigma)}
\end{equation}
where $\sum_{\{\sigma_L\}}$ denotes a sum over all $2^{\vert\Lambda_L\vert}$ spin configurations within $\Lambda_L$. The usual procedure is to solve for 
$\rho^{(L)}$ in some $\Lambda_L$ with $L$ large, followed by taking some infinite sequence of volumes $\Lambda_L$ with $L\to\infty$. The question then 
arises, under what conditions do the thermodynamic quantities under investigation converge in the thermodynamic limit?

The question of convergence (not depending on the choice of sequence of volumes) typically doesn't arise in theoretical physics (outside of mathematical 
physics), because it usually doesn't present a problem. In both the EA and SK spin glasses, most global quantities, such as those mentioned in the Introduction, 
can be shown to converge~\cite{GT02,Guerra03,Talagrand03a} (with probability one in the coupling realizations\footnote{Throughout this paper, all 
conclusions drawn about spin glasses should be understood as occurring with probability one in the coupling realizations unless otherwise noted.}). But if 
we're interested in studying the thermodynamic states themselves (i.e., as mentioned above, the set of all finite-spin correlation functions), then a problem 
may arise at low temperatures, particularly if many pure states are present.

This was addressed in~\cite{NS92}, which considered the EA~Hamiltonian~(\ref{eq:EA}) in an infinite sequence of volumes $\Lambda_L$. The focus there 
was on ``gauge-related'' boundary conditions, i.e., boundary conditions related by a gauge transformation, such as periodic and antiperiodic or any two fixed 
boundary conditions (for a definition and detailed discussion of gauge-related boundary conditions, see Sect.~IX of~\cite{NS22}). Several theorems were proved 
in~\cite{NS92}, but both their statements and proofs will be omitted here. For our purposes, the consequences of the theorems can be summarized as follows. 

Consider an infinite sequence of volumes with periodic boundary conditions as described above, and suppose that there exists only a single pair of globally 
spin-flip related pure states (at positive temperature) or ground states (at zero temperature), such as would occur in either the scaling-droplet 
(SD)~\cite{Mac84,BM85,FH86,FH88b} or trivial--non-trivial (TNT)~\cite{KM00,PY00} pictures.  Then any infinite sequence, chosen independently of 
the couplings, will converge to a 
limiting thermodynamic state: namely, a mixed state comprising the two globally reversed pure/ground states, each with probability 1/2.

Suppose, on the other hand, that the Hamiltonian~(\ref{eq:EA}) supports infinitely many incongruent~\cite{HF87} (i.e., differing by relative interfaces with 
dimensionality equal to the space dimension) pure/ground states, as predicted by the RSB~\cite{Parisi79,Parisi83,MPSTV84a,MPSTV84b,MPV87,MPRRZ00,Read14} 
or chaotic pairs (CP)~\cite{NS96b,NS97,NS98} pictures. Then an infinite sequence of volumes chosen independently of the couplings will generally {\it not\/} 
converge to an infinite-volume (pure or mixed) thermodynamic state, a phenomenon referred to as {\it chaotic size dependence\/}  (CSD)~\cite{NS92,NS03b}. 
Instead, there are infinitely many convergent {\it sub\/}sequences of volumes, each of which converges to a thermodynamic state different from the others. 
Although compactness of the space of spin configurations ensures \cite{chung_book} that the thermodynamic states of appropriately chosen subsequences 
of volumes will converge, in general these subsequences must be chosen dependent on the coupling realization.

Before proceeding, it may be worthwhile to step back and note that nonconvergence of the above type can occur in any system, if the boundary conditions 
are not chosen appropriately. Consider for example a simple uniform Ising ferromagnet in zero field in two or higher dimensions. An infinite sequence of 
volumes with periodic or free boundary conditions will of course converge to a mixture of the positive and negatively magnetized states, each with probability 
$1/2$. But suppose one chose instead fixed random boundary conditions, where for every volume the spins at the boundary are chosen to be $+1$ or $-1$ 
independently in accordance with a fair coin toss. In that case the sequence will not converge, in the following sense: consider any fixed volume $\Lambda_{W}$, 
and consider the spin configuration (at zero temperature) or correlation functions (at positive temperature), within this fixed volume, which we hereafter refer 
to as a ``window''. Then at low temperature, for a sequence of $\Lambda_L$'s with fixed random boundary conditions, the thermodynamic state within the
window $\Lambda_{W}$ will continually flip between the positively and negatively magnetized states (depending on whether the boundary condition on 
$\partial\Lambda_L$ has an excess of plus or minus spins) as $L\to\infty$, never tending toward a limit. In this case there are two subsequences of boundary 
conditions leading to separate limits, one being the positively magnetized and the other the negatively magnetized state.

Of course for the uniform ferromagnet we understand the nature of the broken symmetry and the order parameter, and how to choose appropriate boundary 
conditions to arrive at the desired thermodynamic limit. For the spin glass and similar systems with quenched disorder, however, the situation is far more 
complicated: even knowing the order parameter, whether the EA order parameter~\cite{EA75} or the Parisi order 
parameter~\cite{Parisi79,Parisi83,MPSTV84a,MPSTV84b,MPV87}, or something else, will not solve the problem. We simply don't know how to choose 
coupling-dependent boundary conditions that lead to thermodynamic convergence, and even if we did, these coupling-dependent boundary conditions would  
lead to non-measurability and consequently no clear way of averaging over coupling realizations. Put another way, there is likely no finite procedure for selecting 
convergent subsequences, thereby generating thermodynamic states. CSD arising in this way appears to be a fundamental property of 
Hamiltonians with quenched randomness, and cannot be transformed away by any known methods.\footnote{CSD occurs also in the SK model, 
though in the sense of overlap distributions; see Sect.~5.1 of~\cite{NS03b}.}

The metastate concept was introduced as a tool for handling these sorts of situations; we turn to its definition, construction, and properties in the next section.

\section{Metastates}
\label{sec:meta}

\subsection{Analogy to chaotic dynamical systems}
\label{subsec: chaos}

The term ``chaotic size dependence'' provides more than a picturesque description of the nonconvergence of correlation functions in a coupling-independent 
sequence of volumes: there is a deep analogy to dynamical chaos which provides a clue to resolving the difficulties posed by CSD. Consider the chaotic orbit of 
a particle in a dynamical system: its behavior in time $t$ is deterministic but effectively unpredictable, and can be treated as if it were a random sampling from 
some distribution $\kappa$ on the particle's space of states. One way to construct $\kappa$ is through a histogram that records the proportion of time the particle 
spends in each (coarse-grained) region of states.

Similarly, for systems with quenched disorder such as spin glasses, the behavior of correlation functions as $L$ changes is analogous to the particle's chaotic 
behavior in time $t$. Roughly speaking, the fraction of $\Lambda_L$'s in which a given thermodynamic state appears can be shown to converge as 
$L\to\infty$~\cite{NS96c,NSBerlin,Newman97}, just as the $\kappa$ describing the chaotic behavior of a dynamical system converges as $t\to\infty$. The 
resulting distribution over thermodynamic states carries information on how often a given state appears within $\Lambda_L$'s in the infinite sequence.

Strictly speaking, a thermodynamic state $\Gamma$ is an infinite-volume quantity; by saying that it ``appears'' within a finite volume $\Lambda_L$, we mean 
the following. Fix a window $\Lambda_{W}$ deep inside $\Lambda_L$, i.e., with $W\ll L$; within this window all correlation functions computed using the 
{\it finite-volume\/} Gibbs state $\rho^{(L)}$ are the same as those computed using $\Gamma$ (with negligibly small deviations going to zero as $L\to\infty$ 
with $W$ fixed).  

\subsection{Constructions of metastates}
\label{subsec:construction}

There are several kinds of metastate which carry different kinds of information; here we focus on the simplest variety.  A metastate depends on the Hamiltonian 
(which we will hereafter assume to be~(\ref{eq:EA})), the dimension~$d$, temperature $T$ (which can be zero~\cite{ANSW16} or nonzero~\cite{ANSW14}), 
disorder realization (corresponding here to $J$), and an infinite sequence of volumes $\Lambda_L$ each with a specified boundary condition. We will hereafter 
assume that every $\Lambda_L$ has periodic boundary conditions unless otherwise specified. We will denote a resulting ``periodic boundary condition metastate'' 
by $\kappa_J$, where dependence on all other quantities is suppressed but understood.

There are two independent constructions of metastates, one initially constructed for random-field magnets~\cite{AW90} and one initially constructed for spin 
glasses~\cite{NS96c}. It was proved in~\cite{NSBerlin} (see also~\cite{Newman97}) that there exists an infinite sequence of volumes, chosen independently 
of $J$, for which the two constructions give the same metastate, and so either method can be used, depending on which is more convenient to address the 
problem at hand. Both constructions are sufficiently general that they can be used for a wide variety of applications, including mean-field (MF) Curie-Weiss ferromagnets 
with random couplings~\cite{Kuelske97,Kuelske98}, neural networks~\cite{Kuelske97,vES01}, and other disordered systems.  

The first construction, due to Aizenman and Wehr (AW)~\cite{AW90}, uses a canonical ensemble approach based on varying the couplings {\it outside\/} 
$\Lambda_L$. Let $\rho^{(L)}$ denote the finite-volume Gibbs state (or ground state pair at zero temperature) in $\Lambda_L$ with periodic boundary 
conditions, and consider for each $\Lambda_L$ the random pair $(J_L,\rho^{(L)})$, where $J_L$ is the restriction of $J$ to $\mathbb{E}_L$, and take the 
limit (using compactness \cite{chung_book}) of these finite-dimensional distributions along a $J$-independent subsequence of $L$'s. This yields a probability 
distribution $\kappa$ on infinite-volume pairs $(J,\Gamma)$, where $\Gamma$ denotes a thermodynamic state, which is translation-invariant (because of the 
use of periodic boundary conditions) under simultaneous lattice translations of $J$ and $\rho^{(L)}$. The metastate is then the conditional distribution 
$\kappa_J$ of $\kappa$ given a fixed $J$; it is supported on the infinite-volume thermodynamic states arising from (sub)sequence limits of finite-volume Gibbs 
states.

The second construction, due to Newman and Stein (NS)~\cite{NS96c}, is motivated by the chaotic orbits analogy. Given an infinite sequence of 
volumes $\Lambda_{L_1}$,
$\Lambda_{L_2}$, \ldots with $L_1\ll L_2\ll\ldots$ such that $L_k\to\infty$ as $k\to\infty$, construct a microcanonical ensemble $\kappa_N$ 
in which each of the finite-volume Gibbs states (or ground states at zero temperature) $\rho^{(L_1)},\rho^{(L_2)},\ldots,\rho^{(L_N)}$ has weight 
$N^{-1}$. The ensemble $\kappa_N$ converges to the metastate $\kappa_J$ as $N\to\infty$ (again, possibly with use of a subsequence) in the 
sense that, for every well-behaved function $g(\cdot)$ on thermodynamic states $\Gamma$ (e.g., a function on finitely many spins),
 \begin{equation}
\label{eq:metastate}
\lim_{N\to\infty} N^{-1}\sum_{k=1}^N g(\rho^{(L_k)})=\int g(\Gamma)\ d\kappa_J(\Gamma)\, .
\end{equation}
From~(\ref{eq:metastate}) we see that $\int d\kappa_J(\Gamma)=1$ and $\kappa_J$ can therefore be interpreted as a probability measure on 
thermodynamic states: the finite-volume probability of any event depending on a finite set of spins and/or couplings converges in the infinite-volume 
limit to the $\kappa_J$-probability of that event.

\subsection{Large finite volumes and the thermodynamic limit}
\label{subsec:thermolimit}

A metastate reconciles how nonconvergence in the thermodynamic limit can be used to provide information on the state of a typical 
macroscopically large volume. The information contained in $\kappa_J$ includes the fraction of cube sizes $L_k$ which the system spends in different 
{\it infinite-volume thermodynamic states \/} as $k\to\infty$. If there is only a single pair of pure/ground states, as in 
SD~\cite{Mac84,BM85,FH88b,FH86}, then along any deterministic (i.e., not conditioned on $J$) sequence of volumes the distribution of spin 
configurations in any fixed window $\Lambda_{W}$ generated by the finite-volume Gibbs states will eventually settle down to a fixed state, which as 
discussed above is the restriction of the infinite-volume thermodynamic state to $\Lambda_{W}$. 

On the other hand, if there are many infinite-volume thermodynamic states, as in 
RSB~\cite{Read14,Parisi79,Parisi83,MPSTV84a,MPSTV84b,MPV87,MPRRZ00} (see below), then CSD occurs and the set of correlation 
functions in $\Lambda_{W}$ never converges to a limit. Instead, for any $\Lambda_{L_k}$ with $L_k$ sufficiently large, the set of all 
correlation functions in $\Lambda_{W}$ will be identical to that of one of the many (pure or mixed) infinite-volume thermodynamic states 
available for the system to choose from, and the ``chosen'' state varies with~$L_k$. Although the correlation function values in 
$\Lambda_{W}$ never settle down, the {\it fraction of volumes\/} $\Lambda_{L_k}$ in which a particular thermodynamic states appears in 
$\Lambda_{W}$ {\it does\/} converge (along the subsequence) to a limit, and this information is contained within~$
\kappa_J$~\cite{NSBerlin,Newman97}. 

\subsection{Formal definition and covariance properties of a metastate}
\label{subsec:def}

As mentioned earlier, there are more complex metastate constructions that contain more information than the ``simplest'' metastate discussed above. 
A particularly useful construct at zero temperature is a so-called excitation metastate~\cite{NS2D00,NS2D01,ADNS10,ANS19}), which contains information not 
only on the probability of appearance of different ground states but also all possible local excitations above each ground state. We do not discuss these other 
metastates here but refer the reader to the references for more details.

For the simple periodic boundary condition metastate, we saw in the previous section two different but equivalent constructions. By themselves they do not define 
a metastate, considering that there may be other constructions that also lead to the same result. To arrive at a formal definition, we would like a metastate to satisfy 
certain properties: it must be supported on (infinite-volume) thermodynamic states of the Hamiltonian at positive temperature or ground states at zero temperature, 
and it must be a probability distribution on these thermodynamic states in the sense of~(\ref{eq:metastate}). We would also like a metastate to satisfy certain 
useful covariance properties, to be discussed below.

We begin by formally defining thermodynamic states (i.e.,\ infinite-volume Gibbs states) as probability measures on spin configurations. Let 
$\Sigma=\{-1,+1\}^{\mathbb{Z}^d}$ be the set of all infinite-volume spin configurations and let  $\mathcal M_1(\Sigma)$ be the set of (regular Borel) 
probability measures on $\Sigma$.  An infinite-volume Gibbs state~$\Gamma$ for the  Hamiltonian $H_J$~(\ref{eq:EA}) is an element of $\mathcal 
M_1(\Sigma)$ 
that (at a given temperature $T$) satisfies the Dobrushin-Lanford-Ruelle (DLR) equations~\cite{Georgii88} for that Hamiltonian. Essentially, the DLR equations 
require that the conditional probability in $\Gamma$ of the occurrence of an event in any finite subregion $\Lambda_L$ (in particular, of a given configuration 
of a finite set of spins ${\cal S}\subseteq\Lambda_L$), conditioned on the spins outside $\Lambda_L$, be equal to that obtained from the {\it finite\/}-volume 
Gibbs distribution $\rho^{(L)}$, using the boundary condition on $\partial\Lambda_L$ determined by the spin configuration outside $\Lambda_L$.  
An infinite-volume Gibbs state may be pure or mixed, i.e., a convex combination of two or more pure states. (A pure state is a Gibbs state 
that cannot be expressed as a convex combination of other Gibbs states; see also Sec.\ \ref{subsec:pure}.) We denote the set of infinite-volume Gibbs states 
corresponding to the coupling realization~$J$ by $\mathcal G_J$.

%\begin{df}
%\label{df: metastate}
We can now formally define metastates $\kappa_J$ for the EA Hamiltonian~(\ref{eq:EA}) on $\mathbb{Z}^d$ as follows:

{\bf Definition 3.1} A metastate $\kappa_J$ is a measurable mapping 
\begin{equation}
\begin{aligned}
\mathbb{R}^{\mathbb{E}^d} &\to \mathcal M_1(\Sigma)\\
J &\mapsto \kappa_{J}
\end{aligned}
\end{equation}
with the following properties:

%\medskip 

\begin{enumerate}

\item {\bf Support on Gibbs states.} Every state sampled from $\kappa_J$ is a thermodynamic state for the realization $J$:
\begin{equation}
\kappa_{J}\Bigl(\mathcal G_{J}\Bigr)=1. \nonumber %\ \text{$\nu$-a.s.}\nonumber
\end{equation}

%\smallskip

\item {\bf Coupling Covariance.} For  $B\subset \mathbb{Z}^d$ finite, $J_B\in \mathbb{R}^{\mathbb{E}(B)}$ (the set of edges in $B$), and any 
measurable subset $A$ of  $\mathcal M_1(\Sigma)$,
we define the operation ${\cal L}_{J_B} : \Gamma \mapsto {\cal L}_{J_B}\Gamma$ by its effect on expectation $\langle\cdots\rangle_\Gamma$ 
in $\Gamma$,
\begin{equation}
\label{eq: gamma L}
 \left\langle f(\sigma)\right\rangle_{{\cal L}_{J_B}\Gamma}= \frac{\left\langle f(\sigma) \exp\Bigl(-\beta H_{J_B}(\sigma)\Bigr)\right\rangle_\Gamma}
{\left\langle\exp\Bigl(-\beta H_{J_B}(\sigma)\Bigr)\right\rangle_\Gamma}\, ,
\end{equation}
%Note that the coupling modification is simply a change of density. In particular, it is easy to check that $\Gamma$ and $L_{\Delta J}\Gamma$ 
%are equivalent.
which describes the effect of modifying the couplings within a finite subset $B$ of ${\mathbb Z}^d$. We then require that the metastate 
be covariant under local modifications of the couplings, i.e., 
\begin{equation}
\kappa_{J+J_B}(A)= \kappa_{J}({\cal L}_{J_B}^{-1}A)\nonumber
\end{equation}
where ${\cal L}_{J_B}^{-1} A=\Bigl\{\Gamma\in \mathcal M_1(\Sigma): {\cal L}_{J_B}\Gamma \in A\Bigr\}$.
\end{enumerate}
%\end{df}

We further define a translation-covariant metastate:

{\bf Definition 3.2} A translation-covariant metastate is a metastate $\kappa_J$ with the additional property of 
\begin{enumerate}

\item {\bf Translation Covariance.} For any translation $\tau$ of $\mathbb{Z}^d$ and any measurable subset $A$ of  
$\mathcal M_1(\Sigma)$,
\begin{equation}
\kappa_{\tau J}(A)=\kappa_{J}(\tau^{-1}A).\nonumber
\end{equation}
%\smallskip 

\end{enumerate}

These requirements, including translation covariance, are satisfied by both the AW and NS periodic boundary condition metastate constructions. 
Translation covariance follows easily in the thermodynamic limit from the torus-translation-covariance of periodic boundary conditions. For other 
coupling-independent boundary conditions, metastates can be obtained in a similar way as for periodic boundary conditions, but are not necessarily 
translation covariant; however, translation covariance can be recovered by taking an average of the translates of the finite-volume Gibbs 
measures~\cite{NS2D01,ADNS10}.  In practice, translation covariance is a crucial property without which few conclusions can be drawn.

The importance of coupling covariance, i.e., covariance of a (possibly not translation-covariant) metastate under a local modification of the couplings, 
is less obvious. We note first that under any 
finite change in coupling values, pure states transform to pure states (with some local changes in correlation functions). An important consequence of coupling 
covariance is that if a metastate is supported on some set of pure states, then under any finite coupling transformation $J\to J'$ the transformed metastate 
$\kappa_{J'}$ will be supported on the {\it same\/} set of pure states as $\kappa_J$ (modulo finite changes in certain correlation functions due to the coupling 
transformation).  Coupling covariance is useful because it helps make explicit the dependence of the metastate on any finite set of couplings. This is crucial 
because numerous applications, particularly taking derivatives with respect to coupings, require taking into account the dependence of free energies or other 
thermodynamic quantities on all of the couplings. Moreover, covariance with respect to changes of couplings ensures that taking derivatives with 
respect to couplings will not generate difficult-to-handle boundary terms induced by pure states flowing into or out of a region of integration.

\subsubsection{Gauge invariance of metastates}
\label{subsubsec:gauge}

Definition~3.1, along with the equivalence of AW and NS metastates obtained along certain deterministic sequences of volumes, leads to another covariance 
property of such metastates, proved in~\cite{NS98}, namely that they display gauge invariance with respect to boundary conditions. Recall that constructing a 
metastate requires specification in advance of the boundary conditions on an infinite sequence of volumes. The theorem of~\cite{NS98} shows that a metastate is 
unchanged by {\it any\/} gauge transformation along all or any subset of boundaries $\partial\Lambda_L$ in the sequence. In other words periodic and 
antiperiodic boundary condition metastates are identical, as are any metastates constructed with any combination of periodic and antiperiodic boundary conditions 
within or between volumes. Similarly, all fixed boundary condition metastates are equivalent, for any way of determining the fixed boundary conditions, 
so long as it is independent of $J$. 

This leads to several powerful conclusions; among them is that our restriction in this contribution to periodic boundary condition metastates incurs little loss 
of generality. What {\it is\/} important is that Definition~3.1 almost certainly requires the use of {\it coupling-independent\/} boundary conditions to generate 
metastates. The use of coupling-{\it dependent\/} boundary conditions, though potentially useful for some purposes, would almost certainly result in losing the 
essential properties of measurability, translation-covariance, and coupling-covariance.

\medskip

Now that metastates have been defined and their basic properties enumerated, we can turn to some applications. Our main interest in this contribution is how 
the metastate construct can be used to specify what RSB can (and cannot) mean in the EA model in finite dimensions. We turn to this question in the next section.

%%%%%%%%%%%%%%%%%%%%%%%%%%%%%%%%%%%%%
\section{Metastates and pure states}
\label{sec:mps}

Next we will make some further definitions, and then describe distinct classes of scenarios that are 
possible within the metastate framework. The entire discussion in this section is for infinite-volume systems.

%%%%%%%%%%%%%%%%%%%%%%%
\subsection{Trivial versus non-trivial metastates and the metastate-average state (barycenter)}

We begin by defining what we will mean by a ``trivial metastate''. We say a metastate is trivial if $\kappa_J$
consists of a single atom, in other words a point mass or $\delta$-function on a single Gibbs (i.e.\ DLR) state; 
otherwise, it is nontrivial or ``dispersed''.

An additional construction based on the metastate will be used. Given a metastate, the Gibbs states can be 
averaged using the metastate, to produce the metastate-average state (MAS) or barycenter $\rho_J(\sigma)$ 
\cite{AW90,NS96c,NS97,NS03b}, 
which is itself a Gibbs state (though a rather special one), and which still depends on the sample of $J$. That is, 
$\rho_J(\sigma)=[\Gamma(\sigma)]_{\kappa_J}$, where the square bracket $[\cdots ]_{\kappa_J}$ denotes metastate average, 
in other words
\be
\rho_J(\sigma)=\int \Gamma(\sigma)d\kappa_J(\Gamma).
\ee
If the metastate is trivial, then the MAS $\rho_J=\Gamma$, the Gibbs state on which $\kappa_J$ is concentrated.

%%%%%%%%%%%%%%%%%%%%%%%%%%
\subsection{Pure-state decomposition of a Gibbs state}
\label{subsec:pure}

A convex combination (or mixture) of distinct Gibbs states for the same $H_J$ is again a Gibbs state; 
in general, a convex combination could involve an average taken using a probability measure on Gibbs states. A Gibbs state that 
cannot be expressed as such a combination of other Gibbs states is called an extremal or pure (Gibbs) state. Any Gibbs state can be 
decomposed uniquely into a mixture of pure states for the same $H$ \cite{Georgii88,simon_book}, in the form
\be
\Gamma(\sigma)=\sum_\alpha w_\Gamma(\alpha)\Gamma_\alpha(\sigma),
\label{puredec}
\ee 
where $\Gamma_\alpha$ are pure states, and for any given $\Gamma$ the weights $w_\Gamma(\alpha)\geq 0$ sum to $1$. 
[In practice the decomposition might be continuous and the sum over $\alpha$ would be replaced with an integral using the measure
$w_\Gamma(\alpha)$, but we will usually not show this explicitly.]
Pure states can also be characterized in other, more intrinsic, ways; one of these is that they exhibit clustering of correlations  
\cite{Georgii88,simon_book}. For example, a simple consequence of clustering is that if the Gibbs state $\Gamma$ is pure then
\be
\langle \sigma_{\bf x} (\tau^{-1}\sigma)_{\bf x}\rangle_\Gamma\to \langle\sigma_{\bf x}\rangle_\Gamma\langle(\tau^{-1}\sigma)_{\bf x}\rangle_\Gamma
\ee
as $|{\bf x}-\tau({\bf x})|\to\infty$ [again, $(\tau^{-1}\sigma)_{\bf x}=\sigma_{\tau({\bf x})}$ is a translation of $\sigma_{\bf x}$], 
and a similar statement holds for any two local functions $A(\sigma)$ and $B(\sigma)$ built from spins near ${\bf x}$ taking 
the place of  the two $\sigma_{\bf x}$'s.
The pure states correspond to the ``ordered'' states, which should be familiar in cases such as the uniform Ising ferromagnet
in zero magnetic field, which at low temperature (for dimension $\geq2$) has (at least) two ordered states;
in one of these, the majority of spins in any finite region in a typical configuration are $+1$, while in the other, they are $-1$.  

The Hamiltonian $H_J$ for a classical spin system with quenched disorder may possess a nontrivial group of global ``internal'' 
symmetry operations that act locally on all spins simultaneously, and leave $H_J$ invariant. For example, the EA Hamiltonian in 
eq.\ (\ref{eq:EA}) is invariant under a global spin flip which maps $\sigma_{\bf x}\to -\sigma_{\bf x}$ for all ${\bf x}$. If a magnetic field 
were included, by adding $-h\sum_{\bf x}\sigma_{\bf x}$ to $H_J$, this symmetry would be lost.  From here on we will assume that a metastate 
is obtained using a boundary condition that respects the internal symmetry of the Hamiltonian, in which case the metastate constructions preserve any such 
symmetries:  a Gibbs state drawn from a metastate $\kappa_J$ will be invariant under the full symmetry group of the Hamiltonian, which in turn implies 
symmetries (Ward identities) of correlation functions in that Gibbs state.

In the high-temperature region, there is a unique Gibbs state, which is pure, and the metastate is unique and trivial. 
In this case, the pure state is invariant under the symmetry operations. In a low-temperature phase, it may be that a given pure state
is not invariant under the full symmetry group (when the group is nontrivial), in which case it is said to exhibit spontaneous 
symmetry breaking  \cite{Georgii88,simon_book}. In general, the pure states can be partitioned into {\em orbits}, defined such that 
the symmetry group acts transitively on each orbit. Each orbit consists either of a single invariant pure state, or of more than one pure state. 
As a Gibbs state drawn from the metastate $\kappa_J$ must be invariant under the symmetry, the pure-state decomposition of a 
Gibbs state drawn from $\kappa_J$ must consist of symmetry orbits. We will call a Gibbs state {\em trivial} if its pure-state
decomposition consists of a single symmetry orbit, and {\em nontrivial} if it consists of more than one. (While spontaneous symmetry breaking 
is certainly a nontrivial phenomenon, and is expected to occur at low temperature in many spin-glass systems with nontrivial symmetry groups, this 
terminology will be useful for our analysis of spin-glass phases.) Thus this definition using symmetry orbits can handle all cases, including 
high and low temperature, both discrete and continuous symmetries, and zero and nonzero magnetic field. 

For the MAS, we also define its pure-state decomposition
\bea
\rho_J(\sigma)&=&\sum_\Gamma\kappa_J(\Gamma)\sum_{\alpha\in\Gamma} w_\alpha (\Gamma)\Gamma_\alpha(\sigma)\\
&=&\sum_\alpha \mu(\alpha) \Gamma_\alpha(\sigma),
\eea
where the weights are $\mu(\alpha)=\sum_\Gamma \kappa_J(\Gamma) w_\Gamma(\alpha)=[w_\Gamma(\alpha)]_{\kappa_J}$, 
and implicitly depend on $J$. Here again 
we have written sums, though some of these may in fact be integrals over a probability measure. The same distinction between trivial 
and nontrivial decompositions of a Gibbs state can be applied to this one. 

%%%%%%%%%%%%%%%%%%%%%%%%%%
\subsection{Classes of scenarios for spin glasses}

The preceding definitions of trivial and nontrivial, both for metastates and for Gibbs states, now lead to four combinations 
\cite{NS96c,NS97,Read14}
that provide a broad-brush classification of scenarios for short-range spin glasses, such as the EA model. To simplify the description, we 
assume the Gibbs states have the same character (i.e., either trivial or non-trivial) for almost every Gibbs state drawn from the metastate
(it is unknown whether this must be true, however, the same conclusions would hold if we replaced ``almost surely non-trivial Gibbs 
states'' with ``nonzero probability of non-trivial Gibbs states''). As far as is known to the authors at the time of writing, all four classes 
remain open as possibilities that may occur in some model systems, possibly in different dimensions of space, or at different parameter 
values, for example different temperatures or magnetic fields. They are:
\newline
1) Both the metastate and the Gibbs states are trivial. This is the case in the important SD scenario \cite{Mac84,BM85,FH86,FH88b}, 
as assumed either implicitly \cite{Mac84,BM85}, or (for the Gibbs states) explicitly, following some discussion \cite{FH88b}. Scenarios 
in this class still allow for rich physical phenomena at low temperature \cite{FH88b}. It is also the case that occurs at high temperature.
\newline
2) Nontrivial metastate and trivial Gibbs states. NS \cite{NS96b,NS96c,NS97} termed this class chaotic pairs (CPs), referring to the Ising EA model
at zero magnetic field (the pairs being the symmetry orbits), while the term ``chaotic singles'' has been used for the same but 
with a magnetic field \cite{Read14}. 
\newline
3) Trivial metastate and nontrivial Gibbs states. There is no widely accepted name for this class of scenarios, but below we will mention
its connection with the literature. (NS \cite{NS96b,NS96c,NS97} originally termed one scenario in this class the {\it standard SK picture\/}, 
but we discuss this point in the following subsection.)  
%Many conclusions for this class also apply to the MAS, but note, however, that the transformations 
%involved in coupling covariance were applied to the original $\Gamma$s, not to the MAS.
%A non-trivial MAS can be viewed as belonging to this class, because the underlying metastate, 
%if non-trivial, has merged with the Gibbs states; its MAS is then itself. 
\newline
4) Both the metastate and the Gibbs states are nontrivial. This class is the broadest, and its instances have the richest structure. 
Below we review arguments that the scenario predicted by RSB \cite{Parisi79,Parisi83,MPSTV84a,MPSTV84b,MPV87,MPRRZ00,Read14} 
is in this class in general. (NS \cite{NS96b,NS96c,NS97} originally referred to a scenario in this class as {\it non-standard SK\/}.)
\newline
It follows from this classification that in all classes, except class 1), the MAS is a nontrivial Gibbs state.

Although we mainly discuss positive temperature, it may be useful to say something about $T=0$ also. If the probability distribution 
of the bonds $J$ is continuous without atoms (so {\em not}, for example, the bimodal model, in which 
$J_{{\bf xy}}=\pm J_0$ for $J_0>0$ constant), as assumed so far and as we will continue to assume in this paragraph, then at zero temperature 
a pure state is a ground state (with probability one), that is a single spin configuration. For classical spins, 
ground states always form non-trivial symmetry orbits. Some reflection on the metastate constructions shows that at $T=0$ the Gibbs 
states drawn from the metastate are trivial with probability one. Hence, at $T=0$, only classes 1) and 2) can occur. We may expect 
that if 3) or 4) occur at $T>0$, then as $T\to0$, assuming that no other phase transition intervenes, a scenario in class 3) will reduce 
to one in 1), and one in 4) to one in 2).

%%%%%%%%%%%%%%%%%%%%%%%%%%%
\subsection{Constraints on the scenarios}

Next we discuss some rigorous constraints on the behavior of the scenarios within the preceding classification. 

%%%%%%%%%%%%%%%%%%%%
\subsubsection{Non-self-averaging of overlap distributions}
\label{subsubsec:NSA}

Here we will need to refer to the overlaps of pure states, and to a version~\cite{NS97} of Parisi's overlap distribution function $P(q)$ 
\cite{Parisi83} for the EA model.
Given two pure states $\Gamma_\alpha$, $\Gamma_{\alpha'}$, we can define the average overlap $q_{\alpha\alpha'}$ in the window
$\Lambda_W$, and take $W\to\infty$:
\be
q_{\alpha\alpha'}=\lim_{W\to\infty}\frac{1}{W^d}\sum_{x\in \Lambda_W}\langle \sigma_{\bf x}\rangle_\alpha\langle
\sigma_{\bf x}\rangle_{\alpha'}.
\label{overlaps}
\ee
When $(J,\Gamma)$ are drawn from the distribution $\kappa$, which is translation invariant, and $\Gamma_\alpha$, $\Gamma_{\alpha'}$ 
are drawn from the distribution $w_\Gamma(\alpha) w_\Gamma(\alpha')$ on pairs of pure states in $\Gamma$ [and note that 
$\kappa(J,\Gamma) w_\Gamma(\alpha) w_\Gamma(\alpha')$ is translation invariant], then the ergodic theorem (for translations) implies 
that the $W\to\infty$ limit of the random variable $q_{\alpha\alpha'}$ exists almost surely; the limit is translation invariant. 
Similarly to Parisi \cite{Parisi83}, we then define the probability distribution of overlaps for $(J,\Gamma)$ by
\be
P_{J,\Gamma}(q) =\sum_{\alpha,\alpha'}w_\Gamma(\alpha) w_\Gamma(\alpha') \delta(q-q_{\alpha\alpha'}).
%\label{eq:ovdist}
\ee
We define $P_J(q)=[P_{J,\Gamma}(q)]_{\kappa_J}$ to be the metastate average of $P_{J,\Gamma}(q)$ over $\Gamma$; 
thus if the metastate is trivial, $P_J(q)=P_{J,\Gamma}(q)$. Finally, writing $[\cdots]_\nu$ for expectation using the 
distribution $\nu(J)$ on $J$, we define $P(q)=[P_J(q)]_\nu$. For the MAS $\rho_J$, we define 
$P_{J,\rho}(q)$ similarly to $P_{J,\Gamma}$, using the pure-state decomposition of $\rho_J$ introduced earlier. Each of these distributions
is invariant under translations of $(J,\Gamma)$.
We note that for the EA model at zero magnetic field, the overlaps for the two members of a non-trivial symmetry orbit 
with another fixed pure state have opposite signs, and then $P_{J,\Gamma}(q)$ is an even function of $q$.

A feature of the RSB  MF theory obtained for the infinite-range SK model is that, while the thermodynamic functions
such as the free energy per spin self-average (that is, its thermodynamic limit exists for given disorder, and is almost surely
a constant without fluctuations due to the disorder) \cite{GT02,Guerra03,Talagrand03a}, the same is not true for the distribution 
of overlaps in a given sample of the SK model \cite{ybm,MPSTV84b}. In contrast, for short-range models such as the EA model, 
both $P_J(q)$ and $P_{J,\rho}(q)$ must self average, as a consequence of translation invariance and the 
ergodicity of the distribution $\nu(J)$ \cite{NS96b}. 
Consequently, the picture of RSB in short-range models as belonging to class 3), as seemed to be assumed within the standard 
interpretations of that time, was shown to be inconsistent with the behavior that would be expected on the basis of the RSB 
MF theory. Given the ergodicity of $\nu$, what was called non-self-averaging (NSA) behavior is permitted in a short-range 
model only for $P_{J,\Gamma}(q)$ and only if $\kappa$ is not ergodic, and hence only if $\kappa_J$ is non-trivial \cite{NS96c,NS97}. 
Further, within the RSB scheme (discussed further below), NSA behavior can arise only within class 4), with both a non-trivial metastate 
and non-trivial Gibbs states \cite{Read14}. 

%%%%%%%%%%%%%%%%%%%%%
\subsubsection{Cardinality and structure of pure-state decompositions}

The results of the preceding subsection can be refined further. Using both coupling covariance and translation covariance, 
it was shown in Ref.\ \cite{NS06b} that if the Gibbs states drawn from the metastate are nontrivial, then the pure-state decomposition
of the MAS must be uncountable, forming a continuum without any atoms. Hence for scenarios in class 4), such as RSB,
if the Gibbs states have a decomposition into a countable number of symmetry orbits, as suggested by RSB, then the metastate 
not only must be non-trivial but also must be a continuous atomless distribution on Gibbs states. Also, in class 3) the nontrivial Gibbs 
state must have a decomposition into a continuous atomless distribution on pure states. On the other hand, for scenarios in class 2), 
such as CPs, the metastate is permitted to have support on a countable or even a finite number ($>1$) of trivial Gibbs states. 
We suspect that here too for spin glasses the metastate must be supported on infinitely many such Gibbs states, but as of now 
the question remains open.

%%%%%%%%%%%%%%%%%%%%%
\subsubsection{Complexity of Gibbs states and metastates}

A further refinement invokes the idea of {\it complexity\/} of a pure-state decomposition.
In the form suggested by Palmer \cite{Palmer82}, complexity is the entropy of the set of weights $w_\Gamma(\alpha)$
occurring in the decomposition of a given $\Gamma$; crudely, this corresponds to the logarithm
of the number of pure states, at least if they all have approximately the same weight. This notion
of entropy is well defined, but less useful if it turns out to be infinite. In the latter case, Palmer suggested
calculating it in finite size, and that it might even be extensive, increasing proportionally to the volume
as the system size increases. But this runs into the problem that, in finite size, 
pure states and hence the decomposition are not well defined. Ignoring that difficulty for a moment, it was recognized that 
the complexity cannot be extensive \cite{vEvH84}. Working with well-defined pure states in infinite size, the idea for a given Gibbs state
of counting the number of pure states that can be distinguished when examining the spin configuration
in a finite window was proposed \cite{NS96c,NS97,NS03b}. For a (hyper-)cubic window $\Lambda=\Lambda_W$ of side $W$, 
centered say at the origin, it was expected in a short-range model that the logarithm of the number of distinguishable pure states could 
not grow with $W$ faster than $W^{d-1}$ \cite{NS96c,NS97,NS03b}. For Ising spins with nearest-neighbor interactions, this arises 
from counting the number of boundary conditions on the hypercube that could produce different ground states, and the expectation 
that the same bound would apply for $T>0$.

All these issues can be dealt with in a well-defined way by using some information theory \cite{ct_book}. A pure state
corresponds to a boundary condition, effectively at infinity. The weights and the pure states allow us to consider
the joint distribution $w_\Gamma(\alpha)\Gamma_\alpha(\sigma)$ of pure states $\alpha$ and spin configurations $\sigma$
(for given $\Gamma$). If we restrict $\sigma$ to the spin configuration $\sigma_\Lambda$ in 
the hypercube $\Lambda=\Lambda_W$ as before, then the mutual information $I(\sigma_\Lambda;\alpha)\geq 0$ between 
$\sigma_\Lambda$ and $\alpha$ can be defined in terms of the joint distribution of those random variables, and we now call
this $\Lambda_W$-dependent quantity the {\it complexity\/} $K_\Gamma(\Lambda_W)$ of the Gibbs state $\Gamma$ \cite{hr,Read22}:
\be
K_\Gamma(\Lambda_W)\equiv I_\Gamma(\sigma_\Lambda;\alpha)=\sum_{\sigma_\Lambda,\alpha}w_\Gamma(\alpha)
\Gamma_\alpha(\sigma_\Lambda)\ln \frac{w_\Gamma(\alpha)\Gamma_\alpha(\sigma_\Lambda)}{w_\Gamma(\alpha)
\Gamma(\sigma_\Lambda)}
\ee
[recall that $\Gamma(\sigma_\Lambda)=\sum_\alpha w_\Gamma(\alpha)\Gamma_\alpha(\sigma_\Lambda)$].
This definition does not use any notion of distinguishability of pure states within a window. As $W\to\infty$, $K_\Gamma(\Lambda_W)$ 
increases monotonically,
and tends to the same value as for Palmer's definition of complexity. We can similarly define the complexity $K_{\rho_J}(\Lambda_W)$ 
of the MAS $\rho_J$ in place of $\Gamma$, and also the complexity $K_{\kappa_J}(\Lambda_W)$ of the metastate $\kappa_J$ itself: 
$K_{\kappa_J}(\Lambda_W)$ is defined as the mutual information between the Gibbs (not pure) state $\Gamma$ and 
$\sigma_\Lambda$, using the joint distribution $\kappa_J(\Gamma)\Gamma(\sigma_\Lambda)$ \cite{Read22}. These three quantities 
are related (for given $J$ and $\Lambda_W$) by
\be
K_{\rho_J}(\Lambda_W)=\left[K_\Gamma(\Lambda_W)\right]_{\kappa_J}+K_{\kappa_J}(\Lambda_W).
\ee
For any of these complexities, if it (or its expectation using $\kappa_J$ or $\kappa$) diverges as $W\to\infty$, then we can look for 
the manner in which it diverges. If the leading behavior is $\sim W^{d-\zeta'}$, then $d-\zeta'>0$ indicates an uncountable number 
of pure states (or Gibbs states in the case of $K_{\kappa_J}$). These definitions help to quantify the structures in the above classification,
and the exponents $d-\zeta'$ are expected to be universal within a phase, for example, independent of temperature. 

In the nearest-neighbor EA model of Ising spins in eq.\ (\ref{eq:EA}), a simple argument, similar to counting boundary conditions, 
then produces a bound on the $\kappa$-expectation of the complexity $K_\Gamma(\Lambda_W)$ of the Gibbs states by a constant times 
the surface area $W^{d-1}$ \cite{hr}. More generally, a bound of the same form can be obtained for short-range models of classical spins 
\cite{Read22}, though the argument was valid only for $T>0$. 
(A corresponding result for long-range models was also obtained.) The same bounds also apply to the expectations of the 
other two complexities; thus $\zeta'\geq 1$. Within the usual interpretation of RSB theory, the expected complexity of a Gibbs state 
as $W\to\infty$ is $\lim_{W\to\infty}\left[\left[K_\Gamma(\Lambda_W)\right]_{\kappa_J}\right]_\nu=\psi(1)-\psi(1-x_1)$ \cite{GM84}, 
where $\psi$ is the digamma function and $1-x_1$ is the weight in the $\delta$-function in $P(q)$ at $q(1)$, interpreted as $1-x_1=
\left[\left[\sum_\alpha w_\Gamma(\alpha)^2\right]_{\kappa_J}\right]_\nu\leq 1$.

%%%%%%%%%%%%%%%%%%%
\subsection{Correlations in the MAS}

Like other states, the MAS can be characterized by its spin correlations, and it will be useful for what follows to introduce one.
Due to the quenched disorder, correlation functions in a spin glass will be random with zero mean, and the basic quantities are expectations
of squared correlations.  
%and $\langle\cdots\rangle_\Gamma$ for thermal average in a state $\Gamma$ drawn from 
%the metastate $\kappa_J$, 
Thus we define for Ising spins \cite{Read14,hr}
\bea
C_{\rm MAS}({\bf x},{\bf y})&=&\left[\left(\langle \sigma_{\bf x}\sigma_{\bf y}\rangle_{\rho_J}
-\langle \sigma_{\bf x}\rangle_{\rho_J}\langle\sigma_{\bf y}\rangle_{\rho_J}\right)^2\right]_{\nu}\\
&=&\left[\left(\left[\langle \sigma_{\bf x}\sigma_{\bf y}\rangle_\Gamma\right]_{\kappa_J}
-\left[\langle \sigma_{\bf x}\rangle_\Gamma\right]_{\kappa_J}\left[\langle\sigma_{\bf y}\rangle_\Gamma\right]_{\kappa_J}\right)^2\right]_{\nu}
%C_{\rm MAS}(x,y)&=&[\langle \sigma_x\sigma_y\rangle^2_{\rho_J}]_\nu\\
%&=&\left[\left[\langle \sigma_x\sigma_y\rangle_\Gamma\right]^2_{\kappa_J}\right]_{\nu}.
\eea
using the definition of $\rho_J$. When spin-flip symmetry is present this simplifies, because then $\langle\sigma_{\bf x}\rangle_\Gamma=0$ 
for all ${\bf x}$. 
%For nonzero field we should instead use
%\be
%C_{\rm MAS}(x,y)=\left[\left[\langle \sigma_x\sigma_y\rangle_\Gamma
%-\langle \sigma_x\rangle_\Gamma\langle\sigma_y\rangle_\Gamma\right]^2_{\kappa_J}\right]_{\nu}.
%\ee
This expression resembles the basic correlation function $\chi$ for a spin glass, that is, taking account of the metastate, 
\be
\chi({\bf x},{\bf y})=\left[\left[\left(\langle \sigma_{\bf x}\sigma_{\bf y}\rangle_\Gamma
-\langle \sigma_{\bf x}\rangle_\Gamma\langle\sigma_{\bf y}\rangle_\Gamma\right)^2\right]_{\kappa_J}\right]_{\nu},
\ee
but $C_{\rm MAS}$ differs in that the square is taken {\em after} the metastate averages.
If the metastate is trivial, this of course makes no difference, so $C_{\rm MAS}=\chi$. In the case with spin-flip symmetry, 
$\chi({\bf x},{\bf y})\geq C_{\rm MAS}({\bf x},{\bf y})$. The translation invariance
of $\kappa$ implies that these correlation functions depend on ${\bf x}-{\bf y}$, not ${\bf x}$, ${\bf y}$ separately.
% [the correlation function 
%$\langle \sigma_x\sigma_y\rangle_\Gamma-\langle \sigma_x\rangle_\Gamma\langle\sigma_y\rangle_\Gamma$].

In a spin glass phase, if the decomposition of $\Gamma$ contains more than one pure state (due to either spontaneous symmetry 
breaking or $\Gamma$ being nontrivial), then in most scenarios (at $T>0$) $\chi$ tends to a non-zero constant as $|{\bf x}-{\bf y}|\to\infty$. For a 
non-trivial metastate, $C_{\rm MAS}({\bf x},{\bf y})$ may decay to values $\leq \chi({\bf x},{\bf y})$, especially in those cases in which 
the MAS has an uncountable pure-state decomposition. There are models \cite{NS94,wf,JR10} in which $C_{\rm MAS}$ tends to zero and the 
leading asymptotic behavior is a power law:
\be
C_{\rm MAS}({\bf x}-{\bf y})\sim |{\bf x}-{\bf y}|^{-(d-\zeta)}
\ee
(times a non-universal constant in general) as $|{\bf x}-{\bf y}|\to\infty$, and $\zeta$ is known. 
In the scenarios of class 3), $\rho_J=\Gamma$, and 
$\chi=C_{\rm MAS}$ itself may behave in the fashion described here. In general, when the power law form holds, we expect $d-\zeta$ to 
be a universal constant within a phase. Heuristic arguments in the EA model, to be discussed in the 
following section, further support parts of this picture.  The relation $\zeta=\zeta'$ was proposed, based on those models 
and a fractal picture of ground states \cite{Read14}, but it is not known if it must always hold. 

%%%%%%%%%%%%%%%%%%%%%%%%%%%
\section{Metastate interpretation of replica symmetry breaking}
\label{sec:rsb}

In this Section we explicitly connect RSB and metastate concepts and results. The final arguments are at a theoretical physics 
level of rigor that involve the use of RSB MF theory and fluctuations within replica field theory in short-range models.

We begin with a technical point: even when the magnetic field is nominally zero, the development of RSB appears to
assume that an ``ordering field'' is present, that is a small magnetic field that 
is taken to zero as the system size tends to infinity, but sufficiently slowly that
spin-flip symmetry of the EA model is absent in the states in the limit. It then turns
out that the overlaps $q_{\alpha\alpha'}$ inferred from the RSB MF theory
are almost surely nonnegative. In the following, we will assume this holds.

%%%%%%%%%%%%%%%%%%%
\subsection{$q(x)$ function}

We can define a function $q(x)$ from first principles, motivated by the interpretation of RSB results 
\cite{Parisi83,MPSTV84b,MPV87,Read14}. From $P(q)=P_J(q)$
we first define the cumulative distribution $x(q)=\int_{0^-}^q P(q')\,dq'$ of $q$ as a function on the support of $P(q)$.
On the graph of $x(q)$ versus $q$ we now interchange the axes to produce the graph of a function $q(x)$ for $x\in(0,1)$, 
with the convention that a jump in $x(q)$ becomes a constant region of $q(x)$, while a constant region of $x(q)$ becomes 
a jump discontinuity of $q(x)$; $q(x)$ is a monotonically-increasing function for $x\in(0,1)$. [The value assigned to $q(x)$ 
at a discontinuity in the open interval $(0,1)$ has no known significance.]  We further define $q(1)=\left[\left[\sum_\alpha 
w_\Gamma(\alpha)q_{\alpha\alpha}\right]_{\kappa_J}\right]_\nu=\left[\sum_\alpha \mu(\alpha)\langle 
\sigma_{\bf x}\rangle^2_\alpha\right]_\nu$ (for any ${\bf x}$), where the self-overlap $q_{\alpha\alpha}$, or equivalently the average 
over positions of the time autocorrelation of single spins \cite{EA75}, is the EA order parameter $q_{\rm EA}$ in the pure 
state $\Gamma_\alpha$. (We have recently shown that $q_{\rm EA}$ must be the same for all pure states in the decomposition 
of a given $\Gamma$ drawn from the metastate \cite{NRS}, but whether it is the same for all $\Gamma$ is not known for short-range systems.) 
Similarly, motivated by RSB 
arguments \cite{Read14} given below, we define $q(0)=\left[\sum_{\alpha,\alpha'} \mu(\alpha)\mu(\alpha')q_{\alpha\alpha'}\right]_\nu
=\left[\langle \sigma_{\bf x}\rangle^2_{\rho_J}\right]_\nu$. Then $q(x)$ is now defined on $[0,1]$, and may be discontinuous 
at $x=0$ or $1$. For $q(1)$, this can occur if the measure $w_\Gamma(\alpha)$ is atomless (i.e.\ continuous) for 
a set of $(J,\Gamma)$ of nonzero $\kappa$ probability; for those $\Gamma$, the probability 
$w_\Gamma(\alpha)^2$ of drawing a pure state $\Gamma_\alpha$ twice from $\Gamma$ is zero, and so the $\nu(J)\mu(\alpha)$-expected self-overlap 
$q(1)$ could differ from (in particular, exceed) the supremum of the support of $P(q)$. For $q(0)$, the pairwise overlaps in 
pure states in the MAS could differ from those in the pure states in a single $\Gamma$, with nonzero $\kappa$ probability, 
and so $q(0)$ could differ from (even lie below) the infimum of the support of $P(q)$. It is not clear whether $q(x)$
as we have defined it must necessarily be monotonic for all $x\in[0,1]$; in any case, it contains more information than $P(q)$ does.

We remark that, defining $\overline{q}=\int_0^1dx\,q(x)=\int_{0^-}^{1^+}dq'\,P(q')q'$, 
$q(x)$ obeys $q(1)-\overline{q}\geq 0$ and $\overline{q}-q(0)\geq0$, which follow from the definitions (both quantities are 
sums of variances). If $q(1)-\overline{q}>0$ then the Gibbs states drawn from the metastate are nontrivial, while if 
$\overline{q}-q(0)>0$ then the metastate is nontrivial. If $q(x)$ were monotonic on $[0,1]$, as it is in RSB theory, then both of these 
would hold whenever $q(x)$ is not constant on $(0,1)$ \cite{Read14}. The Cauchy-Schwarz inequality yields $q_{\alpha\alpha'}\leq 
q_{\alpha\alpha}^{1/2}q_{\alpha'\alpha'}^{1/2}$, which would give monotonicity at $x=1$ if $q_{\alpha\alpha}$ were 
independent of $\alpha$, so that $q_{\alpha\alpha}=q_{\alpha'\alpha'}$ for all $\alpha$, $\alpha'$. 

Next we briefly review the basics of RSB (for more detail, see Refs.\ \cite{Parisi79,Parisi83,MPSTV84b,MPV87}). At MF level, the 
free energy per spin has to be obtained by maximizing a functional of the symmetric matrix $Q_{ab}$, 
where $a$, $b$ index the ``replicas'' and run from $1$ to $n$, $Q_{aa}=0$ for all $a$, and the 
functional must be evaluated in the limit $n\to0$ before maximizing. (The maximization rather than the usual minimization 
is a feature of replicas, and is due to the $n\to0$ limit.) Parisi's hierarchical scheme \cite{Parisi79} for breaking the symmetry $S_n$
of the functional under permutations of the replicas is to first divide $n$ into $n/m_1$ blocks of size $m_1$, and assign the value
$q_0$ to elements in the off-diagonal blocks of $Q_{ab}$. Each diagonal block is then divided into sub-blocks of size $m_2$ in the same
fashion (the same for each one), with the value $q_1$ assigned to its off-diagonal sub-blocks. This procedure is iterated
at most a countably-infinite number of times. Finally, when $n\to0$, the block sizes are assumed to obey $0=n\leq m_1\leq m_2\leq \cdots\leq 1$,
giving a countable partition of $[0,1]$, and maximization of the functional is performed with respect to all possible such partitions.
At $n=0$, for a partition, $(q_r)_r$ can be viewed as a function $q(x)$, where $q(x)=q_r$ when $m_r<x<m_{r+1}$ (where $m_0=n=0$), 
$q(0)=q_0$, and $q(1)$ is defined as $q_{r_{\rm max}}$ (where $r_{\rm max}$ is the largest $r$) for a finite partition, and as 
$q(1)=\limsup_{r\to\infty}q_r$ for an infinite one. 
$q(x)$ is found to be a non-negative monotonically increasing function of $x\in[0,1]$; in many situations, $q(x)$ is continuous on the 
open interval $(0,1)$. Its interpretation for $x>0$, similar to that above, has long been standard \cite{Parisi83}.

$q(x)$ as obtained from RSB is not always continuous at $x=0$ or $1$, which brings up an important point. The 
Parisi functional essentially consists of multiple integrals over $x$'s of (products 
of) $q(x)$, so the maximization over functions $q(x)$ is unaffected by changes in the value of $q(x)$ on sets of $x$ of measure zero; 
hence the values $q(0)$, $q(1)$ do not affect the free energy per spin. Sometimes in a given problem, distinct stationary points $q(x)$ 
are found which differ only in the value of $q(1)$ or $q(0)$, and still have the same MF free energy per spin; we have seen that the 
values of these are important for the interpretation. 

As particular examples, in class 1), SD, $q(x)$ would be constant. In the scenarios of class 2), such as CPs, which with an ordering 
field reduce to chaotic singles, pairwise- and self-overlaps of pure states in the MAS should be different, while $P(q)$ consists of a 
single $\delta$-function at $q_{\rm EA}$, and then $q(0)\neq q(x)=q(1)$ for $x>0$. Similarly, scenarios in class 3) can have 
$q(1)\neq q(x)=q(0)$ for $x< 1$ (see the next paragraph). In class 4), there are many possible forms, including $q(x)$ continuous 
but not constant on $[0,1]$. Thus examples in all four classes of scenarios can be characterized by a $q(x)$ function, and examples 
even of classes 2) \cite{hr_unpub} and 3) (see below) can be found at MF level in RSB. However, in the absence of rigorous results 
stating that $q(x)$ is monotonic on $[0,1]$, it is not possible to say that if $q(x)$ is non-constant on $(0,1)$ then both the Gibbs states 
and the metastate are non-trivial (as is the case in RSB); apparently $P(q)=P_J(q)$ could be non-trivial (not a single $\delta$-function) 
due to fluctuations in only one of them.

In MF spin glass models lacking inversion symmetry (i.e.\ spin-flip symmetry, in the Ising case), the form of $q(x)$ in class 3) 
just described was found to occur in RSB MF theory for a range of temperatures, and to correspond to a ``dynamically-frozen'' 
phase in a dynamical approach \cite{kirk1,kirk2,kirk3,kirk4}. It was later argued that 
in a short-range spin glass, this phase, which appeared to possess extensive complexity, would be destroyed by entropic effects \cite{ktw,bb},
so that there would be no transition into it on lowering the temperature (there was still supposed to be another transition at lower temperature, which 
was termed a ``random first-order transition'' \cite{ktw,bb}). However, it does not seem clear that 
such effects must completely destroy such a frozen phase; instead they could leave a similar phase, still in class 3), now having
a subextensive complexity that obeys the bounds discussed above \cite{hr,Read22}.

%%%%%%%%%%%%%%%%%%%%
\subsection{Application of RSB to the AW metastate}

Now we turn to the use of RSB to study the AW metastate for the EA model \cite{Read14}. We consider a finite system in the 
hypercube $\Lambda_L$, with free or periodic boundary conditions. A sub-hypercube $\Lambda_R$ will be viewed as the ``inner''
region, while the sites and edges in $\Lambda_L-\Lambda_R$ constitute the ``outer'' region. We will consider copies of the
system that all experience the same disorder (bonds) in the inner region, but some of which
experience different (independently sampled) disorder in the outer region. An expectation over the disorder 
in the outer region corresponds to an AW metastate expectation using $\kappa_J$, while an expectation over disorder in the inner 
region corresponds to the disorder expectation using $\nu$ (when the limits $L\to\infty$ with $R$ fixed, followed by 
$R\to\infty$ are taken after all the expectations). With this we can study different types
of moments of spin correlation functions, with metastate and disorder averages taken at separate stages;
disorder expectations are taken last, after all metastate expectations have been done. 

To carry out the expectations using the replica formalism, we introduce $l$ ``groups'' of $n_k$ copies or replicas of the system
for the $k$th independent sample of disorder in the outer region ($k=1$, \ldots, $l$); replica indices run from $1$ to
$\sum_kn_k=n$, the total number of replicas in all groups; the replica limit $n_k\to0$ for all $k$ must be taken. 
The replicated theory has permutation symmetry $S_{n_1}\times S_{n_2}\times\cdots S_{n_l}$, broken from $S_n$ 
because of the different disorder experienced by replicas in different groups in the outer region. However, locally in the 
inner region the theory has full $S_n$ symmetry, so the effect is that of a symmetry-breaking perturbation that is at infinity 
once the $L$, $R\to\infty$ limits have been taken.

At MF level, applying the hierarchical ansatz, in the outer region the replicas that experience distinct disorder should have 
small mutual overlaps, but otherwise the structure of the order-parameter matrix $Q_{ab}$ (locally in the outer region) within 
each group will be the same as usual (with $n_k$ in place of $n$). In the inner region, the local $Q_{ab}$ matrix will have the same 
form as when only one group of replicas is used. The presence of the additional partition into blocks of sizes $n_k$ (which can be taken 
all equal) which $\to0$ does not affect the free-energy functional, and the overlaps between replicas in distinct groups will here 
be $q_0$, reflecting the explicit breaking of the $S_n$ symmetry by the outer region, with subsequent further breaking as 
in the usual scheme. 

As a simple example, we first consider the disorder and metastate expectation 
$\left[\left[\langle \sigma_{\bf x}\rangle_\Gamma^2\right]_{\kappa_J}\right]_\nu$ for ${\bf x}$ at say the origin. For this, we choose
distinct replicas $a$, $b$ in say the first group, and because of the distinct stationary points of the MF theory that differ
by permutations in $S_{n_1}$, we should average over those choices; we obtain
\bea
\left[\left[\langle \sigma_{\bf x}\rangle_\Gamma^2\right]_{\kappa_J}\right]_\nu&=&
\lim_{n_1\to 0}\frac{1}{n_1(n_1-1)}\sum_{a, b=1}^{n_1} Q_{ab}\\
&=&\int_0^1 dx\, q(x)
=\overline{q}.
\eea
On the other hand, if we take the metastate expectation before squaring, we must take two replicas from distinct groups,
and then averaging within each group has no effect; we obtain
\be
\left[\left[\langle \sigma_{\bf x}\rangle_\Gamma\right]^2_{\kappa_J}\right]_\nu = q_0=q(0),
\ee
the result claimed earlier \cite{Read14}. 

Similar results are found for more complicated moments. If the metastate and disorder expectations are taken together, the results
are the same as in the old RSB forms. But when some algebraic operation such as a square occurs between the expectations, the 
results differ. In particular, the so-called non-self-averaging (NSA) of $P_{J,\Gamma}(q)$ was found in RSB by using the former type of expectations; 
it occurs whenever $q(x)$ is not constant on $(0,1)$, so NSA occurs in RSB only if both the metastate and the typical Gibbs states are 
non-trivial. When the present formalism is applied to moments of metastate averages, no NSA is found for $P_J(q)$, in complete 
agreement with the considerations of NS \cite{NS96b,NS96c,NS97}. Thus within RSB, fluctuations of $P_{J,\Gamma}(q)$ are 
associated with fluctuations of (nontrivial) $\Gamma$s due to a nontrivial metastate, not the disorder distribution $\nu$ on 
$J$ \cite{NS97,Read14}.

Next we turn to correlation functions. For the correlations in which the metastate and disorder expectations are performed 
together, for example for $\chi({\bf x},{\bf y})$, the results take the traditional form; the replicas used are in the same group. On the other hand, 
for the MAS correlation $C_{\rm MAS}({\bf x},{\bf y})$ one must choose two replicas from distinct groups. In this case it is known in RSB field theory 
\cite{dkt,ddg_book} (though not applied there to the MAS) that the correlation function in the spin glass phase of the EA model at zero magnetic field has 
the asymptotic form
\be
C_{\rm MAS}({\bf x},{\bf y})\sim|{\bf x}-{\bf y}|^{-(d-4)}
\ee
(within a constant factor), so that $\zeta=4$ \cite{Read14}. The calculation here, within a statistical field theory at lowest order 
without any loop corrections, should be valid for dimensions $d>6$. It can be extended to obtain the distribution $P_{J,\rho,W}(q)$ 
of window overlaps, with the overlaps defined as in (\ref{overlaps}) but with $W$ kept fixed, for the MAS $\rho_J$. In leading order 
as $W\to\infty$, again for the EA model in zero magnetic field, the disorder expectation $\left[P_{J,\rho,W}(q)\right]_\nu$ of this 
distribution is found to be a Gaussian of variance $\sim W^{-(d-4)}$ for $d>6$, and further the distribution  $P_{J,\rho,W}(q)$ is found 
to equal its disorder expectation at leading order as $W\to\infty$ \cite{Read14}. Thus in the limit $W\to\infty$, $P_{J,\rho}(q)$ is 
found to be a $\delta$-function at $q=q(0)$, and it self-averages, as expected from Refs.\ \cite{NS96b,NS96c,NS97}.

Some authors carried out a direct numerical study of the AW MAS along these lines, in the three-dimensional 
nearest-neighbor EA model at moderate sizes, and found evidence of a non-trivial metastate \cite{billoire}. Another study 
looked at a dynamical analog \cite{wf} in a one-dimensional power-law model, and found quantitative agreement with the 
exponent that corresponds to  $\zeta=4$ in the short-range case for $d>6$ \cite{wy}; that work has been extended further \cite{jry}.

%%%%%%%%%%%%%%%%%%%%%%%%%%%%%%%%%%%
\section{Conclusion}

In spin glass theory there are outstanding controversies surrounding the nature of the 
spin glass phase: is it a single (pair of) ordered state(s), as suggested by the scaling-droplet theory,
or are there many ordered states, as suggested by replica-symmetry breaking (RSB) mean-field theory? 
The notion of distinct ordered (or ``pure'') states, and hence the question itself, is not even well-defined 
except in an infinite system; hence some sort of infinite-size limit must be taken. Due to the possible presence of 
chaotic size dependence, the limit is not straightforward. The metastate constructions provide a way out of this 
problem. A metastate is an additional layer of structure in the theoretical framework: it is a probability distribution 
on Gibbs states for given disorder, and each Gibbs state could be a mixture of many pure states. A metastate 
contains information about how the states in finite size vary with size or depend on the disorder far from the origin, 
asymptotically at large sizes. Within this framework, a number of significant results that constrain the allowed 
scenarios for a spin glass have been obtained, and some of these results were reviewed in this contribution. 

RSB theory initially takes the form of a mean-field theory for thermodynamic properties. 
When applied to short-range spin glasses, it makes a number of remarkable predictions about the pure states
and their dependence on the sample of disorder. It turns out, at least from a heuristic or physical point of view,
that RSB, when properly interpreted, has so far passed all the consistency tests available from metastate theory, and at least for 
some cases makes predictions about the metastate itself. 

The ultimate answers to the controversies of short-range spin glasses are still unknown.
The nature of the problem raises severe difficulties for traditional analytical methods of theoretical physics,
while numerical methods are subject to the limitations of finite size. It may be that rigorously-proved 
theorems will 
%turn out to 
play a definitive role in resolving the remaining controversies and uncovering 
the true behavior of these fascinating systems.
 
 \acknowledgements
 
 NR is grateful for the support of NSF grant no.\ DMR-1724923.

%\begin{references}
%\end{references}

\bibliography{refs.bib}

\end{document}